\documentclass[final,5p,times,twocolumn,sort&compress]{elsarticle} 

\usepackage{amssymb}
\usepackage{amsmath}
\usepackage{bm}

\usepackage{color}

\journal{Optics Communications}

\begin{document}

\begin{frontmatter}



\title{Macroscopic quantum information processing using spin coherent states}


\author[nii]{Tim Byrnes\corref{cor1}}
\ead{tbyrnes@nii.ac.jp}
\ead[url]{http://nii.timbyrnes.net}
\cortext[cor1]{Corresponding author: +81 3 4212 2000}
\author[nii]{Daniel Rosseau}
\author[mpi,saar]{Megha Khosla}
\author[ipcp]{Alexey Pyrkov}
\author[nii]{Andreas Thomasen}
\author[ntt]{Tetsuya Mukai}
\author[ism]{Shinsuke Koyama}
\author[berk]{Ahmed Abdelrahman}
\author[nii,imo]{Ebubechukwu Ilo-Okeke}
\address[nii]{National Institute of Informatics, 2-1-2 Hitotsubashi, Chiyoda-ku, Tokyo 101-8430, Japan}
\address[mpi]{Max Planck Institute for Informatics, Saarbr{\"u}cken, Germany}
\address[saar]{Saarland University, Saarbr{\"u}cken, Germany}
\address[ntt]{NTT Basic Research Laboratories, NTT Corporation, 3-1 Morinosato-Wakamiya, Atsugi, Kanagawa 243-0198, Japan.}
\address[berk]{Department of Physics, University of California, Berkeley, California 94720, USA.}
\address[ipcp]{Institute of Problems of Chemical Physics RAS, Acad. Semenov av., 1, Chernogolovka, 142432, Russia}
\address[ism]{The Institute of Statistical Mathematics, 10-3 Midori-cho, Tachikawa, Tokyo 190-8562, Japan}
\address[imo]{Department of Physics, School of Science, Federal University of Technology,  P. M. B. 1526, Owerri, Imo State 460001, Nigeria.}

\begin{abstract}
Previously a new scheme of quantum information processing based on spin coherent states of two component Bose-Einstein condensates was
proposed (Byrnes {\it et al.} Phys. Rev. A 85, 40306(R)).  In this paper we give a more detailed exposition of the scheme, expanding on several aspects that were not discussed in full previously.  The basic concept of the scheme is that spin coherent states are used instead of qubits to encode qubit information, and manipulated using collective spin operators.  The scheme goes beyond the continuous variable regime such that the full space of the Bloch sphere is used.  We construct a general framework for quantum algorithms to be executed using multiple spin coherent states, which are individually controlled.  We illustrate the scheme by applications to quantum information protocols, and discuss possible experimental implementations.  Decoherence effects are analyzed under both general conditions and for the experimental implementation proposed. 
\end{abstract}

\begin{keyword}
Bose-Einstein condensates \sep quantum information \sep spin coherent states \sep quantum computing


\PACS 03.75.Kk \sep 03.67.Ac \sep 03.67.Dd


\end{keyword}

\end{frontmatter}


\section{Introduction}
\label{sec:intro}

Bose-Einstein condensation was first achieved in 1995 for ultracold atoms \cite{anderson95,anglin02}, as well as a variety of different physical systems, ranging from exciton-polaritons \cite{deng10}, magnons \cite{demokritov06}, photons \cite{klaers10}, and superfluid Helium \cite{tilley90}.  For atomic Bose-Einstein condensates (BECs), atom chip technology has made possible the miniaturization of traps on the micrometer scale, allowing for the possibility of the individual formation and control of many BECs \cite{fortagh07}.   Due to the long coherence times of cold atoms, a natural application for such systems is quantum information processing, ranging from such tasks as quantum metrology \cite{sorensen00}, quantum simulation \cite{buluta09}, and quantum computing. 

Recently, two component BECs were realized on atom chips realizing full control on the Bloch sphere and spin squeezing \cite{bohi09,riedel10,treutlein06}. The primary application for such two component BECs is currently thought to be for quantum metrology and chip based clocks.  Here we discuss its applications towards quantum computation.  In particular we review a new approach to quantum information processing based on spin coherent states of two component BECs, originally proposed in Ref.  \cite{byrnes12}. While BECs have been considered for quantum computation in the past in works such as Ref. \cite{hecht04}, the results have shown to be generally been unfavorable for these purposes due to enhanced decoherence effects due to the large number of bosons $ N$ in the BEC. The basic idea of the scheme in Ref. \cite{byrnes12} is to take advantage of the analogous state structure of spin coherent states on the Bloch sphere as qubits.  The state of a qubit at a particular location on the Bloch sphere is encoded as a spin coherent state with the same parameters on the Bloch sphere.  Manipulations of the state then proceed by applying collective spin operators $ S^{x,y,z} $ and the entangling operations $ S^z S^z $.  
Using this particular encoding of the quantum information, largely mitigates the problem of decoherence as found in Ref. \cite{hecht04}. We develop the framework for quantum computation using this encoding, illustrated with several quantum algorithms. We also analyze the effects of decoherence from several standpoints and discuss the scheme's performance under a variety of conditions. 

\section{Encoding a single qubit on a spin coherent state}

To encode a qubit, we will consider BECs with ground state degrees of freedom, such as two hyperfine levels in an atomic BEC 
\cite{sorensen00}. We assume that temperatures are sufficiently low such that the spatial degrees of freedom are frozen out. Denote the bosonic annihilation operators of the two ground states as $ a $ and $ b $.  These obey standard bosonic commutation relations $ [a,a^\dagger]= [b,b^\dagger]= 1 $ \cite{li09}. We then propose that a standard qubit state $ \alpha | 0 \rangle + \beta |1 \rangle $ is now encoded on the BEC in the spin coherent state such that
\begin{align}
\label{singlequbitstate}
|\alpha, \beta \rangle \rangle \equiv \frac{1}{\sqrt{N!}} \left( \alpha a^\dagger + \beta b^\dagger \right)^N |0 \rangle ,
\end{align}
where $ \alpha $ and $ \beta $ are arbitrary complex numbers satisfying $ |\alpha |^2 + | \beta |^2 = 1 $.  
Double brackets are used to denote spin coherent states, emphasizing the fact that these are macroscopic states involving many particles.  We call the state (\ref{singlequbitstate}) a ``BEC qubit'' due to the analogous properties of this state with a standard qubit.  For simplicity we consider the boson number $ N = a^\dagger a + b^\dagger b $ to be a conserved number, which amounts to a zero temperature approximation. Assuming $N $ particles that can be in either level $ a $ or $ b $, the Hilbert space has a dimension of $ N + 1 $.  Fock states can be written as
\begin{align}
|k \rangle \equiv \frac{(a^\dagger)^k (b^\dagger)^{N-k}}{\sqrt{k!(N-k)!}} |0 \rangle ,
\label{kdef}
\end{align}
which are orthonormal $ \langle k  |k' \rangle = \delta_{k k'} $ with $ k \in [0,N] $.    

The spin coherent state (\ref{singlequbitstate}) can be visualized by a point on the Bloch sphere with an angular representation 
$ \alpha = \cos(\theta/2) , \beta=\sin(\theta/2) e^{i\phi} $.  The spin coherent states form a set of pseudo-orthogonal states for large $ N $.  The overlap between two states can be calculated to be
\begin{align}
\langle \langle \alpha', \beta' | \alpha, \beta \rangle \rangle   &  = e^{-i (\phi-\phi')N/2} \Big[ \cos \left( \frac{\theta-\theta'}{2} \right) \cos \left(\frac{\phi-\phi'}{2} \right) \nonumber \\
& + i  \cos \left( \frac{\theta+\theta'}{2} \right) \sin \left(\frac{\phi-\phi'}{2} \right) \Big]^N .
\end{align}
For example, for $ \phi=\phi' $ the overlap simplifies to 
\begin{align}
\langle \langle \alpha', \beta' | \alpha, \beta \rangle \rangle = \cos^N \left( \frac{\theta-\theta'}{2} \right) \approx \exp \left( -\frac{N(\theta-\theta')^2}{8} \right) .
\end{align}
Thus beyond angle differences of the order of $ \theta-\theta' \sim 1/\sqrt{N} $, the overlap is exponentially suppressed. 

The state (\ref{singlequbitstate}) can be manipulated using total spin (Schwinger boson) operators 
\begin{align}
S^x & = a^\dagger b + b^\dagger a , \nonumber \\
S^y & = -i a^\dagger b + i b^\dagger a , \nonumber \\
S^z & = a^\dagger a - b^\dagger b ,
\end{align}
which satisfy the usual spin commutation relations $ [S^i,S^j] = 2i \epsilon_{ijk} S^k $, where 
$ \epsilon_{ijk} $ is the Levi-Civita antisymmetric tensor.  In the spin language, (\ref{singlequbitstate}) forms a spin-$ N/2 $ representation of the SU(2) group (we omit the factor of $1/2$ in our spin definition for convenience). For the special case of 
$ N = 1 $, the spin operators reduce to Pauli operators
\begin{align}
\sigma^x & = | 1 \rangle \langle 0 | + | 0 \rangle \langle 1 |, \nonumber \\
\sigma^y & = -i| 1 \rangle \langle 0 | + i| 0 \rangle \langle 1 | , \nonumber \\
\sigma^z & = | 1 \rangle \langle 1 | - | 0\rangle \langle 0 | .
\end{align}
When referring to standard qubits, we will use the $ \sigma^{x,y,z} $ notation throughout this paper to differentiate this to the BEC case where we will use $ S^{x,y,z} $. 

Single BEC qubit rotations can be performed in a completely analogous fashion to regular qubits.  For example, rotations around the $z$-axis of the Bloch sphere can be performed by an evolution
\begin{align}
e^{-i \Omega S^z t} |\alpha, \beta \rangle \rangle & = 
\frac{1}{\sqrt{N!}} \sum_{k=0}^N {N \choose k} ( \alpha a^\dagger e^{-i\Omega t})^k ( \beta b^\dagger e^{i \Omega t})^{N-k} |0 \rangle \nonumber \\
& = |\alpha e^{-i \Omega t}, \beta e^{i \Omega t}\rangle \rangle .
\label{rotationz}
\end{align}
Similar rotations may be performed around any axis by an application of
\begin{align}
H_{1} = \hbar \Omega \bm{n} \cdot \bm{S} =  \hbar \Omega ( n_x S^x + n_y S^y +n_z S^z )
\end{align}
where $ \bm{n} = (n_x,n_y,n_z) $ is a unit vector. Expectation values of the total spin are identical to that of a single spin (up to a factor of $ N $), taking values
\begin{align}
\langle S^x \rangle & = N(\alpha^* \beta + \alpha \beta^*) \nonumber \\
\langle S^y \rangle & = N(-i \alpha^* \beta + i \alpha \beta^*) \nonumber \\
\langle S^z \rangle & = N( | \alpha |^2 - | \beta |^2 ) ,
\end{align}
where $ \langle S^{x,y,z} \rangle \equiv \langle \langle \alpha, \beta | S^{x,y,z} | \alpha, \beta  \rangle \rangle $.  
These may be derived efficiently by using the relations
\begin{align}
[S^x , \alpha a^\dagger + \beta b^\dagger ] & = \alpha b^\dagger + \beta a^\dagger  \nonumber \\
[S^y , \alpha a^\dagger + \beta b^\dagger ] & = -i \alpha b^\dagger + i\beta a^\dagger  \nonumber \\
[S^z , \alpha a^\dagger + \beta b^\dagger ] & = \alpha a^\dagger - \beta b^\dagger 
\end{align}
and 
\begin{align}
[\alpha^* a + \beta^* b  , \alpha a^\dagger + \beta b^\dagger ] = 1 .
\end{align}

In contrast to the average spin, when normalized according to $ S^{x,y,z}/N $ has the same result as for standard qubits, 
variance diminishes under the same normalization:
\begin{align}
\label{variance}
\frac{\langle (S^z)^2 \rangle - \langle S^z \rangle^2}{N^2} = \frac{4 |\alpha \beta |^2}{N} .
\end{align}
This is in accordance to widespread notion that for $ N \rightarrow \infty $ the spins approach ``classical'' variables. We shall however see in the following section that despite the classical appearance of such a state, such a many boson state can exhibit quantum properties such as entanglement. 

We note that collective state encodings have been proposed previously in works such as Refs. \cite{lukin01,rabl06,brion07},
where a large number of particles is used to encode a two level system. A key difference between the encoding in these works and (\ref{singlequbitstate}) is that the full $ N+1 $ Hilbert space is used here to encode the two level system. Typically in these works first the spins are polarized in a particular direction and low lying excitations are used to encode quantum information. In contrast, for various parameters $ \alpha,\beta $ the state (\ref{singlequbitstate}) uses the full Hilbert space of the spins.  Thus although many physical particles encode the quantum state, the Hilbert space mapping is one-to-one.

\begin{figure}
\scalebox{0.4}{\includegraphics{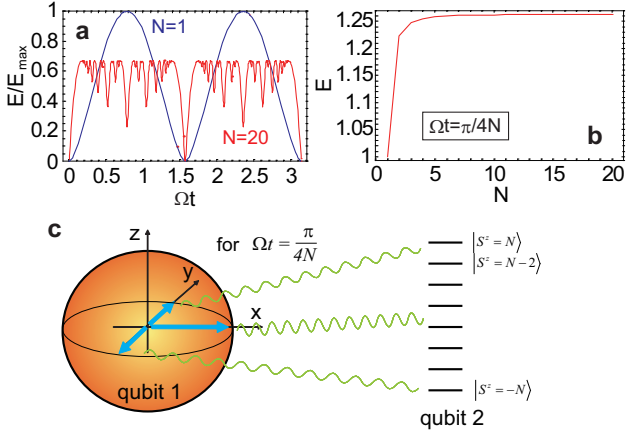}}
\caption{\label{fig2}
{\bf a} The entanglement normalized to the maximum entanglement ($E_{\mbox{\tiny max}} = \log_2 (N+1) $) between two BEC qubits for the particle numbers as shown. {\bf b} Entanglement at a time $ \Omega t= \pi/4N $ for various boson numbers $ N $.  {\bf c} A schematic representation of the entangled state (\ref{bosonqubitentanglement}), reproduced from Ref. \cite{byrnes12}. }
\end{figure}

\section{Entanglement between BECs}

Two BEC qubit gates can be formed by any product of the Schwinger boson operators of the form
\begin{align}
\label{generaltwoqubit}
H_2 = \sum_{n,m=1}^M \sum_{i,j = x,y,z} \hbar \Omega_{i j} S^{i}_{n} S^{j}_m 
\end{align}
where $ \Omega_{i j} $ are real symmetric parameters. Our first aim will be to show that such an operator, combined with 
$ H_1 $ allows for a set of operations with the corresponding operations to standard qubit operations.  To make this definition more 
precise, let us consider the most general Hamiltonian for standard qubits:
\begin{align}
\label{generaltwostandardqubit}
H = \sum_{ \bm{j}}  A(\bm{j}) \prod_{n=1}^M \sigma_n^{j(n)}
\end{align}
where $ \bm{j} $ is a vector of length $ M $ with entries $ j(n) = 0,x,y,z $, and the sum over $ j $ runs over every 
combination.  We assume the convention that $ \sigma^0 = I $ is the identity matrix.  The  $ A(\bm{j}) $ are arbitrary coefficients.  
Accordingly, we will say for the BEC qubit case that we may perform the corresponding operations to standard qubit operations if we can realize the Hamiltonian
\begin{align}
H = \sum_{ \bm{j}}  A(\bm{j}) \prod_{n=1}^M S_n^{j(n)}
\end{align}
where again we assume the convention that $S^0 = I $ is the identity and other definitions are the same as the standard qubit case. 

A well known result from quantum control theory states that if it is possible to perform an operation with Hamiltonians $ H_A $ and $ H_B $, then it is also possible to perform the operation corresponding to $ H_C = i [H_A,H_B] $  \cite{lloyd95}. Therefore, the combination of $ H_1 $ and $ H_2$ may be combined to form an arbitrary Hamiltonian involving spin operators according to universality arguments. It is simple to show that by successive commutations of $ H_1 $ and $ H_2$ an arbitrary product of spin Hamiltonians 
\begin{align}
\label{prodspinham}
H \propto \prod_{n=1}^M S_n^{j(n)}
\end{align}
may be produced.  An arbitrary sum of such Hamiltonian may then be produced for example by a Trotter expansion, which is of the form (\ref{generaltwoqubit}) \cite{lloyd95}. For BEC qubits in general higher order operators may be constructed (e.g. $ S_n^l$ with $ l\ge 2 $). However, our aim here is to produce the corresponding operations to a standard qubit system using the BEC qubits hence are unnecessary for our purposes. 

Such two BEC interactions naturally possess a bosonic enhancement which can result in short gate times. To see this, note that
Pauli operators are $ \sigma^j \sim O(1) $ while the Schwinger boson operators are $ S^j \sim O(N) $.  This makes the two BEC qubit interaction $ H_2 \sim O(N^2) $.  The effect of the boosted energy scale of the interaction can be observed by examining explicitly the state evolution of two BEC qubits. Let us consider henceforth the interaction Hamiltonian 
\begin{align}
\label{zzham}
H_2 = \hbar \Omega S^z_1 S^z_2 .
\end{align}
This may be done without any loss of generality since (\ref{generaltwoqubit}) can be converted to (\ref{zzham}) by universality arguments. 
As a simple illustration of two BEC entanglement, let us perform the analogue of the maximally entangling operation 
\begin{align}
& e^{-i  \sigma^z_1 \sigma^z_2 \frac{\pi}{4} } ( | \uparrow \rangle + | \downarrow \rangle ) ( | \uparrow \rangle + | \downarrow \rangle ) =  | + y \rangle | \uparrow \rangle  + | - y \rangle | \downarrow \rangle, \label{qubitentangler}
\end{align}
where $ | \pm y \rangle = e^{\mp i\frac{\pi}{4}} | \uparrow \rangle + e^{\pm i\frac{\pi}{4}} | \downarrow \rangle $. Starting from two unentangled BEC qubits, we may apply $ H_2 $ to obtain
\begin{align}
&  | \Psi (t) \rangle = e^{-i \Omega S^z_1 S^z_2 t} | \frac{1}{\sqrt{2}}, \frac{1}{\sqrt{2}} \rangle \rangle | \frac{1}{\sqrt{2}}, \frac{1}{\sqrt{2}} \rangle \rangle 
= \nonumber \\
& \frac{1}{\sqrt{2^N}} \sum_{k_2} \sqrt{N \choose k_2}  | \frac{e^{i(N-2 k_2)\Omega t}}{\sqrt{2}}  , \frac{e^{-i(N-2 k_2) \Omega t}}{\sqrt{2}}  \rangle \rangle |k_2 \rangle ,
\label{bosonqubitentanglement}
\end{align}
where we have used the normalized eigenstates of the $ S^z $ operator (\ref{kdef}).  For gate times equal to $ \Omega t = \pi/4N   $ we obtain the analogous state to (\ref{qubitentangler}). For example, the maximum $ S^z $ eigenstates $ |k_2=0,N \rangle $ on BEC qubit 2 are entangled with the states  $ | \pm y \rangle \rangle \equiv | \frac{e^{\pm i\pi/4}}{\sqrt{2}}  , \frac{e^{\mp i \pi/4}}{\sqrt{2}} \rangle \rangle  $, which is the analogue of a Bell state for the BEC qubits. A visualization of the state (\ref{bosonqubitentanglement}) is shown in Figure \ref{fig2}(c). For each $z$-eigenstate on BEC qubit 2, there is a state $ | \frac{e^{i(N-2 k_2) \pi/4N}}{\sqrt{2}}  , \frac{e^{-i(N-2 k_2) \pi/4N}}{\sqrt{2}}  \rangle \rangle  $ on BEC qubit 1 represented on the Bloch sphere entangled with it. The type of entangled state is a continuous version of the original qubit state (\ref{qubitentangler}), and has similarities to continuous variable formulations of quantum computing \cite{braunstein05}, although the class of states that are used here are quite different. 

The effect of the boosted energy scale of (\ref{zzham}) is that a gate time of $\Omega t=\pi/4N  $ was required to produce this entangled state, in comparison to the standard qubit case of $ \Omega t=\pi/4  $.  The origin of the 
reduced gate time is due to the bosonic enhancement of the interaction Hamiltonian, originating from the boosted energy scale of many particles occupying the same quantum state in the BEC.  An example of the speedup for the case of atom chips will be given in the section relating to the experimental implementation. 

Despite the widespread belief that for $ N \rightarrow \infty $ the spins approach classical variables according to (\ref{variance}), the entangling operation (\ref{bosonqubitentanglement}) generates genuine entanglement between the BEC qubits.  As a measure of the entanglement, in Figure \ref{fig2}(a) we plot the von Neumann entropy 
$ E = - \mbox{Tr} ( \rho_1 \log_2 \rho_1 ) $, where $ \rho_1  = \mbox{Tr}_2 | \Psi (t) \rangle \langle \Psi (t)| $ has the partial trace over the degrees of freedom in BEC 2 taken \cite{nielsen00}.  For the standard qubit case ($N=1$), the entropy reaches its maximal value at $ \Omega t = \pi/4  $ in accordance with (\ref{qubitentangler}).  For the BEC qubit case there is an initial sharp rise, corresponding to the improvement in speed of the entangling operation, but later saturates to a non-maximal value due to the presence of the binomial factors in (\ref{bosonqubitentanglement}) biasing the states towards zero spin values. This saturating value approaches $ \lim_{N \rightarrow \infty} E/E_{\mbox{\tiny max}} \approx 1/2 $ \cite{byrnes13}, showing that 
macroscopic entanglement can indeed survive even in the ``classical'' limit of $ N \rightarrow \infty $.   In Figure \ref{fig2}(b) we show the amount of entanglement present at times $\Omega t=\pi/4N  $.  We see that at such times there is approximately the same amount of entanglement as for the $ N= 1 $ case as for large $ N $, confirming that the $ e^{-iS^z_1 S^z_2 \pi/4N} $ gate gives the bosonic analogy to the operation (\ref{qubitentangler}).  

In a realistic experimental situation, it is very challenging to have perfect control of the atom number on each BEC.  In the above discussion we have assumed for simplicity that the atom numbers $ N_1 $ and $ N_2 $ on BEC qubits 1 and 2 respectively are equal $ N_1 = N_2 = N $.  For unequal atom numbers the entangling operation creates the state
\begin{align}
& e^{-i \Omega S^z_1 S^z_2 t} | \frac{1}{\sqrt{2}}, \frac{1}{\sqrt{2}} \rangle \rangle_{N_1} | \frac{1}{\sqrt{2}}, \frac{1}{\sqrt{2}} \rangle \rangle_{N_2} 
= \nonumber \\
& \frac{1}{\sqrt{2^{N_2}}} \sum_{k_2=0}^{N_2} \sqrt{N_2 \choose k_2}  | \frac{e^{i(N_2-2 k_2)\Omega t}}{\sqrt{2}}  , \frac{e^{-i(N_2-2 k_2) \Omega t}}{\sqrt{2}}  \rangle \rangle_{N_1}  |k_2 \rangle =  \label{bah} \\
& \frac{1}{\sqrt{2^{N_1}}} \sum_{k_1=0}^{N_1} \sqrt{N_1 \choose k_1}  |k_1 \rangle  
| \frac{e^{i(N_1-2 k_1)\Omega t}}{\sqrt{2}}  , \frac{e^{-i(N_1-2 k_1) \Omega t}}{\sqrt{2}}  \rangle \rangle_{N_2} .
\label{bosonqubitentanglement2}
\end{align}
We therefore have the same general structure, but a different distribution of spin coherent states according to the phase factors $ e^{\pm i(N_{1,2}-2 k_{1,2})\Omega t} $.  Taking the example of a rotation with $ \Omega t = \pi/4N $ as an example, having fluctuations $ N_{1,2} = N + \delta N_{1,2} $ amounts to having a ``fan'' of  spin coherent states with ends which are not exactly distributed in the $\pm y $ directions in Figure \ref{fig2}(c).  The error of this is $\delta N_{1,2}/N $,  which for the work of Ref. \cite{treutlein08} was estimated to be at a level of $ \sim 4\% $. 
If $ N_{1,2} $ can be measured accurately, this variation can be compensated by choosing 
appropriate gate times $ \Omega t = \pi/4 N_{1,2} $. However, only one of the distributions of the coherent states can be fixed as can be seen from (\ref{bah}) and (\ref{bosonqubitentanglement2}).  While absorption imaging can be done with an accuracy of several atoms \cite{treutlein08}, this destroys the BEC hence non-destructive methods such as phase contrast imaging would be required \cite{ilookeke14}. Thus the effect of number fluctuations can be considered to be an effective gate error, which may be mitigated if the particle numbers are known.  If the precise particle numbers are unknown, this contributes to an effective dephasing on average which acts on the two BEC entangling gate.  One BEC qubit coherent operations are immune to the variation in particle number as may be observed by examining (\ref{rotationz}) where the phases are independent of the particle number $ N $.

\begin{figure}[ht]
\scalebox{0.4}{\includegraphics{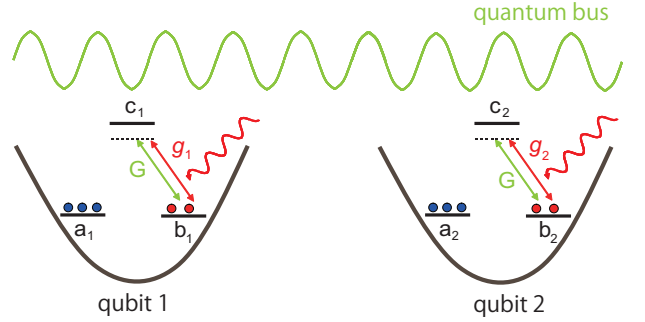}}
\caption{\label{fig1} Two BEC qubits mediated by a quantum bus. The quantum bus couples transitions between levels $ b $ and $ c $ with energy $ G $. Individual pulses coupling levels $ b $ and $ c $ with energy $ g $ create an effective $ S^z S^z $ interaction between the two BECs.  }
\end{figure}

\section{Quantum algorithms}

For a particular quantum algorithm designed on standard qubits, how does one translate this to BEC qubits? Due to the larger Hilbert space of the bosonic system, there is in fact no unique mapping -- there are many different possible solutions. However, there are a few constraints that we demand for a suitable mapping.  These are:
\begin{itemize}
\item A simple mapping between the input and output states of the algorithm on BEC qubits to its qubit counterpart exists 
\item The operations should not involve complex controls beyond linear products of spin operators (\ref{prodspinham})
\item The algorithm should be robust against decoherence (does not involve encodings on states that are sensitive to decoherence, e.g. Schrodinger cat states)
\end{itemize}
The first of these criteria ensures that the BEC qubit version of the algorithm in fact does perform effectively the same quantum computation as its qubit counterpart. Once the quantum algorithm is complete, one should be able to read off the result of quantum computation by a readout of the BEC qubits, which may involve some simple encoding rule to obtain the standard qubit version. As has been discussed in the previous sections, the BEC qubits do not have precisely the same properties as qubits, but they are similar in many respects.  The question is, {\it are they similar enough such that for the purposes of performing a quantum algorithm, they can be used instead of standard qubits}? The hypothesis of quantum computing using spin coherent states is that the answer to this question is ``yes'', although a conclusive answer to this question is still in the process of being investigated.  The second and third criteria ensure that no experimentally unrealistic situations are assumed in constructing the quantum algorithm, as the current scheme aims to produce demonstrable protocols in the lab.  

Finding a quantum algorithm which satisfies the above properties is a nontrivial task, but for many applications a good starting point amounts to: (i) finding the sequence of Hamiltonians required for the algorithm, (ii) making the replacement $ \sigma^j_n \rightarrow N S^j_n $, $ \sigma^i_n \sigma^j_m  \rightarrow S^i_n S^j_m $, (iii) Evolving the same sequence of Hamiltonians for a reduced time $ t \rightarrow t/N $.  This approach is reasonable from the point of view that we are performing the same algorithm except that a higher representation of SU(2) is being used. 

Let us illustrate this procedure with the simple example of Deutsch's algorithm. We reformulate the standard qubit version ($N=1$) of the algorithm in the following form convenient for our purposes \cite{nielsen00}.  The oracle performing the function $ | x \rangle | y \rangle \rightarrow | x \rangle | f(x) \oplus y \rangle $ is assumed to be one of the four Hamiltonians $ H_D = \{ 0, 2\sigma^z_2,\sigma^z_1 \sigma^z_2 + \sigma^z_2 -1,-\sigma^z_1 \sigma^z_2 + \sigma^z_2 -1 \} $ and evolved for a time $ t=\pi/4 $, which correspond to the functions $ f(x) = \{ (0,0),(1,1),(0,1),(1,0) \} $ respectively. The initial state is assumed to be the state $ (\uparrow + \downarrow) \uparrow $, and a measurement of BEC qubit 1 in the $ x$-basis distinguishes between constant and balanced functions via the results $ (\uparrow + \downarrow)$ and $ (\uparrow - \downarrow)$ respectively. 

This can be translated into the corresponding algorithm for BEC qubits according to the following procedure.  The oracle is assumed to be one of the following Hamiltonians $ H_D = \{ 0, 2 N S^z_2,S^z_1 S^z_2 + N S^z_2 -N^2,-S^z_1 S^z_2 + N S^z_2 -N^2 \} $, and we prepare the initial state as $ | \frac{1}{\sqrt{2}} , \frac{1}{\sqrt{2}} \rangle \rangle | 1 , 0 \rangle \rangle $. After evolving the Hamiltonians for a time $ t = \pi/4N $, we obtain (up to an overall phase)
\begin{align}
e^{-i H_D \pi/4N } | \frac{1}{\sqrt{2}} , \frac{1}{\sqrt{2}} \rangle \rangle | 1 , 0 \rangle \rangle = 
| \frac{1}{\sqrt{2}} , \pm \frac{1}{\sqrt{2}} \rangle \rangle | 1 , 0 \rangle \rangle,
\end{align}
where $ + $ is obtained for the constant cases and $ - $ for the balanced cases. A measurement of BEC qubit 1 distinguishes the constant and balanced cases with one evaluation of the oracle, which is precisely the same result as the qubit version of the Deutsch's algorithm.   

Another quantum algorithm for which the mapping has been demonstrated to date is quantum teleportation \cite{pyrkov13b}, where a BEC qubit is transferred between two parties by the use of shared entanglement. 
We refer the reader to Ref. \cite{pyrkov13b} for further details regarding the protocol.  In the case of teleportation the recipe $ \sigma^j_n \rightarrow N S^j_n $, $ \sigma^i_n \sigma^j_m  \rightarrow S^i_n S^j_m $ requires some modification in order to satisfy the expected properties of teleportation.  Specifically, the entangling times that are used are $ \Omega t = 1/\sqrt{2N}$ rather than  $ \Omega t = \pi/4N$ discussed above.

\section{Experimental implementation}
\label{sec:expimp}

The most promising system for realizing the current scheme is using BECs on atom chips, as many BECs may be placed close together, and BEC spin coherent states have been realized and manipulated  \cite{treutlein06,bohi09,riedel10}.  
We now describe the specific experimental configuration for the above theory applying it to this case. In these works, the $ |F=1, m_F=-1 \rangle $ and $ |F=2, m_F=1 \rangle $ hyperfine levels of the $5\mbox{S}_{1/2} $ ground state of $^{87}$Rb are used as the qubit states. In terms of Figure \ref{fig1}, we make the association for the operator $ a^\dagger $ ($ b^\dagger $) as creating an atom in the state $  |F=1, m_F=-1 \rangle $ ($ |F=2, m_F=1 \rangle $).  Since the BEC contains a large number of atoms, there can be more than one atom present in a particular level, as illustrated in Figure \ref{fig1}. Level ``c'' in Figure \ref{fig1} corresponds to a suitable higher energy level satisfying optical selection rules determined by the polarization of the laser fields.  For example, taking the $ b \leftrightarrow c $ transitions to be $ \sigma^- $ circularly polarized light, we make the association that the $ c^\dagger $ operator creates an atom in the state $ | F'=2, m_F'=0 \rangle $ of the $5\mbox{P}_{3/2} $ state.  

Single qubit rotations may be performed according to existing methods using microwave pulses as discussed in Refs. \cite{treutlein06,bohi09}. Here we discuss an alternative all-optical method for single qubit rotations which naturally fits into the scheme for two qubit rotations (discussed below) \cite{abdelrahman14}.  Using detuned pulses we may connect levels $ a $ and $ b $ via an adiabatic passage using the two transitions shown in Figure \ref{fig1}.  These are
\begin{align}
H_1 & = \Delta c^\dagger c + g ( a^\dagger c + c^\dagger a) +  g (b^\dagger c + c^\dagger b) 
\label{singlequbitadiabatic}
\end{align}
Here $ c^\dagger $ is a creation operator for a boson in level $ c $ and $ \Delta $ is the detuning between the laser pulse and the transition energy. Assuming that $ \Delta \gg g $, the effective coupling between levels $a$ and $b$ is then 
\begin{align}
H_1^{\mbox{\tiny eff}} =  \frac{g^2}{\Delta} ( a^\dagger b + b^\dagger a) =  \frac{g^2}{\Delta} S^x.
\end{align}

There is however a complication with a straightforward application of the above scheme, which is that in order to create a transition between $ |F=1, m_F=-1 \rangle $ and $ |F=2, m_F=1 \rangle $ levels, the nuclear spin must be necessarily flipped, which can only occur by the hyperfine coupling \cite{waxman07}. The effective Rabi frequency can be derived in two equivalent ways, using a a two level adiabatic passage as in Ref. \cite{abdelrahman14}, or by considering the interference between hyperfine coupled basis \cite{waxman07}. In either case the effective Rabi frequency is 
\begin{equation}
\hbar \Omega_1^{\mbox{\tiny eff}} = \frac{g^2 \delta E }{\Delta^2}  .
\end{equation}
where $ \delta E $ is the hyperfine splitting of the optically excited states.  $ S^z $ rotations are performed by exploiting the natural energy difference between the $ F=1 $ and $ F=2 $ levels $ \Omega_z/2\pi \sim 6.8 $GHz, which allows for full control of the single qubit state on the Bloch sphere. 

Two qubit gates may be implemented by using a quantum bus \cite{pellizzari95}, which is implemented by connecting two BEC qubits via cavity QED, as shown in Figure \ref{fig1}.  Recent experimental advances have allowed for the possibility of incorporating cavity QED with atom chips \cite{colombe07,purdy08}.  A scheme for entangling two BECs via cavity QED was described in Ref. \cite{pyrkov13}.  
In order to perform the entangling operation (\ref{zzham}), the two BECs corresponding to the two qubits are placed within the cavity, with a resonant frequency detuned off the $ b \leftrightarrow c $ transition as for the single qubit case.  Due to the large detuning, without the presence of the laser induced transition $ b \leftrightarrow c $, the cavity has no effect on the states. The two BEC qubit gate can be turned on and off on demand by the application of the laser connecting levels $ b $ and $ c $. 

To model such a system, consider an interaction Hamiltonian \cite{pyrkov13,rosseau14}
\begin{align}
\label{qubushamiltonian}
H_2 = \frac{\hbar \omega_0}{2} \sum_{n=1,2} F^z_n + \hbar \omega p^\dagger p + G  \sum_{n=1,2} \left[ F^-_n p^\dagger + F^+_n p \right] ,
\end{align}
where $ F^z =c^\dagger c - b^\dagger b$, $ F^+ = c^\dagger b  $, $ \omega_0 $ is the transition frequency, and $ p $ is the photon annihilation operator. Assuming a large detuning $ \Delta = \hbar \omega_0 - \hbar \omega \gg G $, we may adiabatically eliminate the photons and the excited state we obtain an effective Hamiltonian 
\begin{align}
\label{expandedinteraction}
H_2^{\mbox{\tiny eff}} \approx \hbar \Omega_2^{\mbox{\tiny eff}} (2 S^z_1 S^z_2 - (S^z_1)^2 - (S^z_2)^2  )
\end{align}
where first order spin operators have been dropped and
\begin{align}
\hbar \Omega_2^{\mbox{\tiny eff}} =  -\frac{G^2 g^2 }{4 \Delta^3} .
\end{align}
The energy scale of the interaction term is then proportional to $ N^2 $ as claimed previously. Although this interaction involves
undesired single qubit interaction terms $ \sim (S^z)^2 $, these may be eliminated and 
converted to the form  $ \propto S^z_1 S^z_2 $ by either performing a cancellation process as described in \cite{pyrkov13}, or by concatenating this with single qubit gates using universality arguments \cite{lloyd95}.  

Finally, we discuss how measurements can be performed on the BECs. There are primarily two classifications for measurement, which
can be classified according to whether it is destructive or nondestructive with respect to the BEC itself. While the quantum state is always perturbed in a measurement, in a destructive measurement the BEC itself is destroyed in the process, such that it cannot be repeated more than once for a particular experimental instance. These include techniques such as absorption imaging \cite{anderson95,andrews97b} or fluorescent imaging~\cite{depue00} which are examples of strong projective measurements.  In these cases, what is finally performed is a count of the number of atoms in each level, which is a measurement in the $S^z $ basis. On the other hand nondestructive techniques using non-resonant detuned light~\cite{andrews96,bradley97b,higbie05,kohnen11,brahms12,gajdacz13} have been used to measure the properties of ultracold atomic gases~\cite{kohnen11,brahms12,gajdacz13}, as well as small and dense atomic condensates~\cite{andrews96,bradley97b,higbie05}. In particular, in phase contrast imaging (PCI) \cite{andrews96,bradley97b,higbie05} coherent light illuminates the BEC and a state dependent phase shift develops in the light. By measuring the phase shift via interference, the state of the BEC can be inferred. The PCI measurement does not destroy the atomic condensate, it can be applied repeatedly~\cite{andrews97,meppelink10} on the same atomic sample. This also amounts to an estimate of $S^z $, but is not a strong measurement and instead results in some dephasing of the coherence between the levels \cite{ilookeke14}.

\begin{figure}[ht]
\scalebox{0.35}{\includegraphics{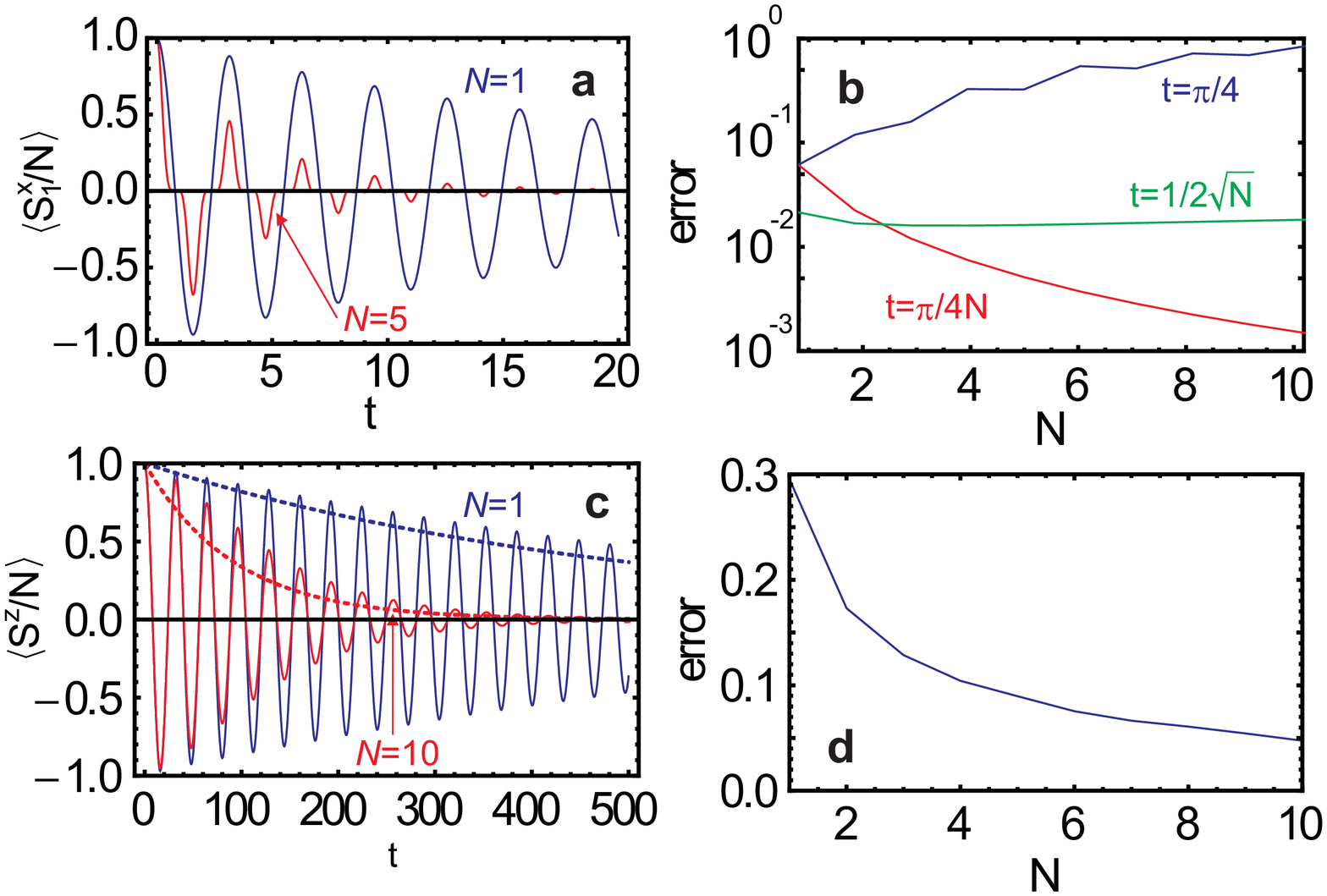}}
\caption{\label{fig4} Decoherence effects of using bosonic qubits for coherent operations.  (a) Evolution under a two qubit  $ S^x_1 S^x_2 $ gate under $ S^z $-type decoherence.  Parameters used are $ \Gamma_z = 0.01 $ and $ \hbar \Omega_2 = 1 $ in (\ref{twoqubitdephasing}).  (b) The error of the two qubit operation as described in Fig. \ref{fig1} as function of boson number $ N $ for various evolution times as shown.  
(c)  Decoherence of Rabi oscillations due to spontaneous emission of an intermediate adiabatic passage state (solid lines).  Parameters used are $ g=1 $, $ \Delta = 10 $, $ \Gamma_s = 0.1 $ in (\ref{spontaneousmaster}).  The effective decay rate $ \exp[-\Gamma_1^{\mbox{\tiny eff}} t] $ is plotted for comparison (dotted lines).  (d) The error dependence with $ N $ for the two qubit gate induced by cavity QED.  Parameters used are $ \Gamma_c = 1 $, $ G= 1 $, $ \Delta = 10 $ in (\ref{cavityphotondecay}). The dotted line shows the effective decay rate $ \exp[-\Gamma_2^{\mbox{\tiny eff}} t] $.}
\end{figure}

\section{Decoherence}

We now consider decoherence effects due to the use of BEC qubits.  We first examine the effects of dephasing and single particle loss in a generic way. In this case special emphasis will be made on the scaling properties of the 
decoherence with $ N $, which is typically a large number in our case.  We later give the various decoherence channels from a physical perspective under various situations.

\subsection{General properties of decoherence on BEC qubits}

\subsubsection{Dephasing and single particle loss for state storage}

The first scenario we consider is when a quantum state is stored in the system of qubits and no gates are applied, i.e. when the BEC qubits are used to simply store a state. Let us assume the two generic channels of decoherence of dephasing and single particle loss. 
Considering dephasing first, we model this via the master equation 
\begin{align}
\frac{d \rho}{dt} = - \frac{\Gamma_z}{2} \sum_{n=1}^M [ (S^z_n)^2 \rho - 2 S^z_n \rho S^z_n + \rho(S^z_n)^2 ],
\end{align}
where $ \Gamma_z $ is the dephasing rate. For a standard qubit register, the 
information in a general quantum state can be reconstructed by $ 4^M - 1 $ expectation values of $ (I_1,\sigma^x_1,\sigma^y_1,\sigma^z_1) \otimes (I_2,\sigma^x_2,\sigma^y_2,\sigma^z_2)  \dots \otimes (I_M,\sigma^x_M,\sigma^y_M,\sigma^z_M) $ \cite{altepeter04}.  For the bosonic system, we can consider the same correlations but with the replacement $ \sigma \rightarrow S $, but there are in general higher order correlations involving 
powers of operators beyond order one, but these are unnecessary for our purposes as previously discussed. 

Examining the dephasing of the general correlation $ \langle \prod_n S^{j(n)}_n \rangle $ where $ j(n) = I,x,y,z $, we
obtain the evolution equation $ d\langle \prod_n S^{j(n)}_n \rangle/dt = -2 \Gamma_z K_z \langle \prod_n S^{j(n)}_n \rangle $, which can be solved to give
\begin{align}
\label{dephasingeqn}
\langle  \prod_n S^{j(n)}_n \rangle \propto \exp[-2 \Gamma_z K_z t] .
\end{align}
Here $ K_z $ is the number of non-commuting $ S^{j(n)}_n $ operators with $ S^{z}_n $ (i.e.  $ j(n) = x,y $), which is  
independent of $ N $ and is at most equal to $ M $.  The crucial aspect to note here is that the above equation does not have any $ N $ dependence. In fact the equation is identical to that for the standard  qubit case ($N=1$).  Physically this difference is due to the statistical independence of the dephasing processes among the bosons. 

For single particle loss, we consider the Hamiltonian
\begin{align}
\frac{d \rho}{dt} = - \frac{\Gamma_l}{2} \sum_{n=1}^M [ a_n^\dagger a_n \rho - 2 a_n \rho a_n^\dagger + \rho a_n^\dagger a_n +b_n^\dagger b_n \rho - 2 b_n \rho b_n^\dagger + \rho b_n^\dagger b_n],
\end{align}
where $ \Gamma_l $ is the particle loss rate. We find the similar result
\begin{align}
\label{particlelosseqn}
\langle  \prod_n S^{j(n)}_n \rangle \propto \exp[- \Gamma_l K_{l} t] ,
\end{align}
where $ K_{l} $ is the number of $ S^{j(n)}_n $ operators that are not the identity (i.e.  $ j(n) = x,y,z $), which is 
again independent of $ N $ and is at most equal to $ M$. The general results of (\ref{dephasingeqn}) and (\ref{particlelosseqn}) show that {\it dephasing and single particle loss is not enhanced by the use of BEC qubits
when they are used to store a spin coherent state}.  For an implementation using atom chip BECs, the dephasing time $ 1/\Gamma_z$ has been estimated to be on the order of seconds \cite{treutlein06}, which is highly competitive in comparison to
other systems proposed for quantum computation \cite{ladd10}.  

The origin of this behavior is that powers of the spin operators beyond one (e.g. $ (S^x_n)^2 $) are not used to encode any quantum information in our scheme. An extreme case that would be highly susceptible to decoherence would be the use of Schrodinger cat states such as $ \alpha | 1,0 \rangle \rangle + \beta | 0, 1 \rangle \rangle $ to encode quantum information \cite{hecht04}.  Such states are highly vulnerable to decoherence, due to the high order spin correlations $ \langle (S^x_n)^N \rangle - \langle S^x_n \rangle^N $ present for such a state. We note that 
although we only considered single particle loss here, in general there exist higher order loss effects such as three-body recombination which would increase with particle number. However, as we show in Sec. \ref{sec:particleloss}, for atom chips these are in practice rather small effects and are not detrimental to the scheme.

\subsubsection{Dephasing under continuous operation}

The results of the previous subsection suggest that as long as we only observe correlations of the form $ \langle  \prod_n S^{j(n)}_n \rangle $, the system is stable against accelerated decoherence effects due to the large boson numbers involved.  Here we show that 
depending on the state trajectory that is traversed by the quantum algorithm, there are circumstances where decoherence effects can be enhanced. We discuss ways to avoid this situation.

The simplest example where this occurs is for the two qubit operation $ H_{2} = \hbar \Omega_2 S^z_1 S^z_2 $ under $S^x$-type dephasing. The 
master equation for this is
\begin{align}
\label{twoqubitdephasing}
\frac{d \rho}{dt} = i [ \rho, H_{2}] - \frac{\Gamma_z}{2} \sum_{n=1}^2 [ (S_n^x)^2 \rho - 2 S_n^x \rho S_n^x + \rho (S_n^x)^2 ] .
\end{align}
In Figure \ref{fig4}a we show results showing the expectation value of $ \langle S_1^x \rangle $ after preparing both the qubits in $ S^x = N $ eigenstates.  We see that there is a degradation of the oscillations with increasing $ N $.  An 
understanding of the origin of this enhanced decoherence can be obtained by examining the structure of the states at particular times.  For example, at $ \Omega_2 t = \pi/4 $, (\ref{bosonqubitentanglement}) can be 
written
\begin{align}
& \frac{1}{2} \left( | \frac{1}{\sqrt{2}} , \frac{1}{\sqrt{2}} \rangle \rangle_1  + | \frac{1}{\sqrt{2}} , -\frac{1}{\sqrt{2}} \rangle \rangle_1  \right) | \frac{e^{i \pi N/4}}{\sqrt{2}} , \frac{e^{-i \pi N/4}}{\sqrt{2}} \rangle \rangle_2 \nonumber \\
& + \frac{1}{2} \left( | \frac{1}{\sqrt{2}} , \frac{1}{\sqrt{2}} \rangle \rangle_1  - | \frac{1}{\sqrt{2}} , -\frac{1}{\sqrt{2}} \rangle \rangle_1  \right) | -\frac{e^{i \pi N/4}}{\sqrt{2}} , \frac{e^{-i \pi N/4}}{\sqrt{2}} \rangle \rangle_2 .
\label{schrodingercat}
\end{align}
which is an entangled Schrodinger cat state.  Thus although the system returns to a product state with period $ \Omega_2 T = \pi/2  $, during its evolution it traverses fragile states that are highly susceptible to decoherence \cite{byrnes13}.  

What does the observation mean for quantum processing using BEC states? Fortunately, as we have discussed in previous sections, in the mapping of standard qubit algorithms to the bosonic qubits, gates producing such highly entangled states often can be avoided during the construction of the algorithm.  For example, the mapping of the CNOT operation requires gate times of $ \Omega_2 t = \pi/4N $.  At such times the state does not involve Schrodinger cat states, as may be observed from (\ref{bosonqubitentanglement}). A demonstration of the error for these short gate times is shown in Figure \ref{fig4}(b). The errors are defined to be the 
value $ 1- \frac{\langle S_1^z \rangle}{N}$ after evolving under $ H_{xx} $ for the times shown, then reversing the operation by application of  $ - H_{xx} $ for an equal time.  We see that as the boson number is increased a monotonic decrease in error is achieved. This can be understood as originating from the fast two qubit gates that are possible using bosons,  under the same dephasing rate allowing for an improved fidelity of operation. For long gate operations such as $ \Omega_2 t = \pi/4 $ we see an increase of the error with $ N $, which is also evident in Figure \ref{fig4}(a).  The critical time beyond which the state is highly susceptible to decoherence appears to be approximately $ \Omega_2 t \sim 1/2\sqrt{N} $, which coincides with the timescale required for entanglement of order $ E \sim O(E_{\mbox{\tiny max}}) $ to occur in Figure \ref{fig1}a.

\subsection{Physical channels of decoherence}

In the previous sections we have attributed the decoherence mechanisms of dephasing and particle loss in a generic way without considering their physical origins.  We henceforth consider more specifically the physical origins of each of these effects. Specifically we consider the specific decoherence channels introduced by the optical manipulation scheme as described in Sec. \ref{sec:expimp}.  These decohering effects occur only during the single and two BEC qubit operations, unlike other effects such as particle loss which are always present.     

\subsubsection{Particle loss in atom chips}
\label{sec:particleloss}

For particle loss in atom chip BECs we have two major contributing factors: interactions with the background and inelastic collisions between the atoms \cite{treutlein08}. 
We can model these factors with the following rate equation
\begin{align}
   \label{particlenumbereqn}
   \frac{1}{N} \frac{dN}{dt}=-\Gamma_{l}-K \left\langle n \right\rangle-L \left\langle n^2 \right\rangle
\end{align}
where $\Gamma_{l}$ is loss due to background interactions, $n$ is the density of particles, $K$ is the two-body scattering rate, and $L$ is the three-body recombination rate, and $ N $ is the atom number. Note that the effects of both two-body scattering and three-body recombination scale with particle density, thus are more significant at high densities.  

As we encode our quantum information using two atomic hyperfine states, for our chosen two states of the BEC the 
loss equation is \cite{treutlein08}
\begin{align}
   \label{1,-1}
   \frac{1}{N_a} \frac{dN_a}{dt}=-\Gamma_{l}-K_{ab} \left\langle n_b \right\rangle-L_a \left\langle n_a^2 \right\rangle
   \\
   \label{2,1}
   \frac{1}{N_b} \frac{dN_b}{dt}=-\Gamma_{l}-K_{ab} \left\langle n_a \right\rangle-K_b \left\langle n_b \right\rangle
\end{align}
here the subscripts label the two hyperfine states $a \equiv \left |F=1,m_F = -1 \right\rangle$ and $b \equiv \left| F=2, m_F = 1 \right\rangle$, and $K_{ab}$ is a rate constant for collision between the two states, and $ N_{a,b} $ are the populations of the respective hyperfine states. There is no two-body scattering term, $K_a $, in (\ref{1,-1}) because such collisions are forbidden due to conservation of energy and angular momentum selection rules \cite{pethick08}. For BECs in standard (non-atom chip) magneto-optical traps, the three-body recombination is typically the dominant effect as shown in \cite{burt97}, with an estimated value of $L_a=5.8(1.9)\times 10^{-30}$cm$^6$ s$^{-1}$. However, this is due to the relatively high density of BECs in this configuration, and in atom chips the densities are much lower making this a less serious effect. The three-body recombination term is left out of (\ref{2,1}) because the two-body term remains and is much more significant, with rate constants $K_b=1.194(19)\times10^{-13}$ cm$^3$ s$^{-1}$ and $K_{ab}=0.780(19)\times10^{-13}$ cm$^3$ s$^{-1}$ as measured in Ref. \cite{mertes07}. Lastly the background loss, has been measured for atomic chips to be no greater than order $\Gamma_{l} = 10^{-1}$ s$^{-1}$ \cite{treutlein08}. 

With these experimental values  we can calculate some estimates for these effects. On atom chips the particle densities being of order $10^{12}$ cm$^{-2}$ \cite{treutlein08}, giving a three body-recombination lifetime of order $10^{6}$ s, and the two-body inelastic collisions in state $\left| 2,1 \right\rangle$, and between states, giving a lifetime of $10$ s, which is comparable to the background loss. Thus the three body-recombination 
is negligible compared to both background loss and two-body scattering. In either case, these long lifetimes mean in practice the loss should not contribute significantly to the decoherence of the BEC qubits.

\subsubsection{Spontaneous emission}

We now turn to additional sources of decoherence that may occur due to the experimental implementation that is used \cite{pyrkov13}.
Specifically, for the single qubit rotations which involve an adiabatic passage through excited states as discussed in the 
previous section, spontaneous emission can occur which can affect the fidelity of the operation.  To model this effect we consider a single $\Lambda $-scheme as shown in Figure \ref{fig1}, and consider the master equation
\begin{align}
\frac{d \rho}{dt} = & i [ \rho,H_1] - \frac{\Gamma_s}{2} \left[ c^\dagger a a^\dagger c \rho - 2 a^\dagger c \rho c^\dagger a +
\rho c^\dagger a a^\dagger c \right]\nonumber \\
& - \frac{\Gamma_s}{2} \left[ c^\dagger b b^\dagger c \rho - 2 b^\dagger c \rho c^\dagger b +
\rho c^\dagger b b^\dagger c \right]
\label{spontaneousmaster}
\end{align}
where $ H_1 $ is given in (\ref{singlequbitadiabatic}).  
Here we have assumed that the coupling $ g $ to the intermediate level from the logical states $ a $ and $ b $ are equal, and $ \Delta $ is the detuning to level $ c $. Equations (\ref{spontaneousmaster}) can be solved for arbitrary $ N $ \cite{pyrkov13}, and we show the results in Figure \ref{fig4}(c). For large detunings $ \Delta \gg g $ we see Rabi oscillations between levels $ a $ and $ b $, resulting in the oscillations corresponding to an effective coupling 
\begin{align}
\hbar \Omega_1^{\mbox{\tiny eff}} = \frac{g^2}{\Delta} .
\label{hbaromega1}
\end{align}
The decoherence is found to have an effective rate of 
\begin{align}
\label{effectiveonequbitdecoherence}
\Gamma_1^{\mbox{\tiny eff}} \approx \frac{g^2 \Gamma_s (N+1)}{\Delta^2} .
\end{align}
The factor of $ N+1 $ originate from the enhanced spontaneous emission due to final state stimulation of the bosons. 
Although (\ref{effectiveonequbitdecoherence}) has a scaling proportional to $ N $ which appears to be detrimental to the scheme, there are several ways to overcome this.  As the detuning $ \Delta $ is a free experimental parameter, this may be chosen to be sufficiently large to overcome the factor of $ N +1 $ in the numerator.  The 
drawback to this is that (\ref{hbaromega1}) also becomes rather small, corresponding to long gate times.  In Ref. \cite{abdelrahman14}, it was shown that by a suitable choice of parameters it is possible to obtain gates in the range of $\sim $ MHz, which still exceed the gate speeds based on microwave pulses.  An alternative method based on the stimulated Raman adiabatic passage (STIRAP) potentially offers a far more superior approach, as it involves dark states not involving the excited state at all \cite{thomasen15}. By eliminating the excited state contribution this greatly suppresses the dephasing due to spontaneous emission, which is one of the main drawbacks of using optical methods to control BECs. Therefore, while (\ref{effectiveonequbitdecoherence}) captures the main disadvantage of optical coherent control of BECs, by the use of suitably designed optical pulses we expect that this is experimentally viable.

\subsubsection{Cavity photon loss}

Another mechanism of decoherence is via cavity photon decay, for the two BEC qubit interactions implemented by the quantum bus methods described in the previous section.  We model the decoherence for this process via the master equation
\begin{align}
\label{cavityphotondecay}
\frac{d \rho}{dt} = i [ \rho,H_{2}] - \frac{\Gamma_c}{2} \left[ p^\dagger p \rho - 2 p^\dagger \rho p +
\rho p^\dagger p \right]
\end{align}
where $ H_{2} $ is given in (\ref{qubushamiltonian}).  We expect that this form of decoherence is independent of $ N $, since the cavity photon decay rate is in general independent on photon population in the cavity. 
Since (\ref{cavityphotondecay}) produces an effective two BEC interaction, it is susceptible to the enhanced
decoherence effects as were discussed in previous sections. 
For this reason we restrict our discussion to short gate times with $ \Omega_2^{\mbox{\tiny eff}} t = \pi/4N $. In Ref. \cite{pyrkov13} 
it was found that the effective of the cavity decay may be summarized by the effective decoherence rate
\begin{align}
\Gamma_2^{\mbox{\tiny eff}} \approx \frac{G^2 \Gamma_c}{\Delta^2} .
\label{cavityphotondecoherence}
\end{align}
which is independent of the boson number,  as long as
we use gates on the timescale $ \Omega_2^{\mbox{\tiny eff}} t = \pi/4N $. The scaling of the error of the two BEC qubit gate as a function of $N $ is shown in Figure \ref{fig4}(d). The figure shows qualitatively the same behavior as Figure \ref{fig4}(b), where the error monotonically decreases with the number of bosons.

\section{Discussion}

We have described a framework for performing quantum information processing that is based on spin coherent states, looking at the specific example of two component spinor BECs. The theory has similarities to the theory of continuous variables quantum information processing \cite{braunstein05}, where instead of using discrete variables, an effectively continuous Hilbert space is used. However, while current uses of spin coherent states in continuous variables uses only small deviations from polarized states on the Bloch sphere, here we enter a fully 
non-linear regime using the full space of the Bloch sphere.  We discussed entanglement properties of entangling two BEC qubits, and mapping procedures for converting standard qubit quantum algorithms to the BEC qubit case.  A specific implementation using atom chips were discussed, together with expected decoherence 
effects associated with this implementation. From a conceptual point of view, one of the interesting results of the proposed scheme is that
that despite the ``classical'' $ N \rightarrow \infty $ limit, entanglement, and hence quantum computation  can be performed in the system when entangling gates of the form $ S_1^i S_2^j $ are applied. This said, depending on the type of state that is targeted, it may be difficult to observe such macroscopic entanglement for the same reason that Schrodinger cat states are difficult to observe, due to enhanced decoherence rates of such states.  However, with a judicious use of quantum states, the approach may offer an alternative to standard quantum computation schemes beyond standard qubit and continuous variable schemes.

\section*{Acknowledgments}

T.B. thanks Daniel Burgarth for comments regarding the manuscript. This work is supported by the Transdisciplinary Research Integration Center, the Okawa Foundation, the Inamori Foundation, NTT Basic Laboratories, and JSPS KAKENHI Grant Number 26790061.


\section*{References}




\end{document}